\documentclass{l4dc2027}

\usepackage{xcolor}
\usepackage{xurl}
\usepackage[moderate]{savetrees}

\newtheorem{ass}{Assumption}
\newtheorem{prb}{Problem}
\newtheorem{rmk}{Remark}

\title[Data-driven control of polynomial systems using implicit methods]{Data-driven control of polynomial systems using implicit methods}
\usepackage{times}

\author{
 \Name{Andr{\'a}s Sasfi} \Email{asasfi@control.ee.ethz.ch}\\
 \Name{Florian D{\"o}rfler} \Email{doerfler@control.ee.ethz.ch}\\
 \addr Department of Information Technology and Electrical Engineering, ETH Z{\"u}rich, Switzerland
 \vspace{-0.3em}\AND
 \Name{Ivan Markovsky} \Email{imarkovsky@cimne.upc.edu}\\
 \addr Catalan Institution for Research and Advanced Studies (ICREA) and International Centre for Numerical Methods in Engineering (CIMNE), Barcelona, Spain
 \vspace{-0.3em}\AND
 \Name{Jaap Eising} \Email{j.eising@rug.nl}\\
 \addr ENTEG, University of Groningen, the Netherlands
}

\begin{document}

\maketitle
\vspace{-0.5em}\begin{abstract}
While conventional behavioral data-driven control represents linear time-invariant (LTI) systems directly from data, extending these data representations to nonlinear systems remains an open challenge.
We address this gap for general polynomial systems, characterizing their finite-length trajectories as a real algebraic variety defined by implicit polynomial relations rather than explicit equations.
First, we provide conditions on measured data under which these data can be used to represent the system. 
Building on this representation, we formulate data-driven simulation and predictive control problems that are equivalent to their model-based counterparts, and solve the resulting polynomial systems globally using homotopy continuation.
Numerical case studies confirm the effectiveness of this approach by successfully simulating a stiff Van der Pol oscillator in situations where explicit models fail. Furthermore, these data-driven simulation and control methods are illustrated on a Duffing oscillator, demonstrating that our method outperforms Koopman-based approaches in predicting long trajectories.
\end{abstract}


\section{Introduction}
Data-driven predictive control based on behavioral systems theory has become popular for linear systems~\citep{coulson2019data,berberich2020data,dorfler2022bridging}.
The behavioral approach describes a system by the set of its trajectories rather than by a particular representation through equations.
For linear time-invariant (LTI) systems, the set of trajectories is a subspace, and this geometric viewpoint allows for a direct data representation of the behavior by the Fundamental Lemma~\citep{willems2005note,markovsky2023identifiability}.
For nonlinear systems, the set of trajectories is not a linear subspace, and extending such direct data representations is an open problem, as also highlighted in the reviews by~\citet{markovsky2021behavioral,berberich2024overview}.
In this work, we focus on polynomial systems, for which this set is a real algebraic variety, and we show that it can be represented directly by data, enabling data-driven simulation and control.

\textbf{Related work:}
in the data-driven control literature, nonlinear systems are often described by explicit models in discrete-time, such as nonlinear state-space or nonlinear ARX (NARX) equations~\citep{martin2023guarantees,de2023learning,bonassi2022offset}.
While powerful, these explicit descriptions treat variables asymmetrically by isolating the future state or output, which often fails to capture the nonlinear geometry of the system.
On the other hand, the behavioral approach preserves the geometric structure by treating all variables symmetrically, enabling implicit descriptions, see Section~\ref{sec:integration} for an illustration.
Data-driven methods inspired by the Fundamental Lemma have also been extended to nonlinear systems~\citep{berberich2024overview}.
Many of these extensions are restricted to specific system classes, such as Volterra~\citep{rueda2020data}, Hammerstein and Wiener~\citep{berberich2020trajectory}, or generalized bilinear systems~\citep{markovsky2022bilinear}, while others address general nonlinear systems through NARX models~\citep{mishra2021NARX,fazzi2024simulation}, basis functions~\citep{lazar2024basis}, kernel methods~\citep{huang2023robust}, or local linearization~\citep{berberich2022linear}.
Most importantly, all these methods rely on explicit equations, which forgoes the main advantage of the geometric, representation-free viewpoint.

Koopman operator methods lift the state by a dictionary of observables and fit a linear model in the lifted space~\citep{williams2015data,korda2018linear,mezic2026tutorial}.
This general framework is suitable for describing implicit relations in the data.
In practice, however, the lifted state is propagated by the linear model and mapped back to the state by a linear decoder~\citep{mezic2026tutorial}, so nonlinear relations between the observables are typically lost along the prediction.
Reprojection onto the set of lifted states restores this consistency~\citep{vangoor2023reprojection,mezic2026tutorial}, but the predictions remain approximate unless the observables span a Koopman-invariant subspace, which can be infinite-dimensional even for polynomial dynamics.

Beyond data-driven control, relevant approaches also exist in the learning and system identification literature.
For instance, the SINDy framework identifies nonlinear dynamics by regressing the time derivative of the state on a library of candidate functions with sparse coefficients~\citep{brunton2016discovering}.
While \citet{mangan2016inferring,kaheman2020sindy} propose implicit variants of the SINDy algorithm, they aim at identifying continuous-time dynamics, and therefore require measuring or estimating time derivatives, which can be restrictive.
This same limitation applies to \citet{guo2022data,huang2025data} that design controllers for continuous-time polynomial systems via sum-of-squares programs.
Finally, manifold learning methods also exploit that the data lie on a low-dimensional manifold, 
learned together with the dynamics on it~\citep{floryan2022data} or via a diffusion model~\citep{kharitenko2026landing}, but these approximations lack conditions on the data that guarantee an exact representation.

\textbf{Contributions:}
our approach overcomes the limitations of the existing works by characterizing implicit relations between the variables.
We represent these relations directly by data, and solve the resulting polynomial equations instead of relying on linear methods.
Specifically, our contributions are as follows. 
\begin{itemize}
    \item We motivate implicit data-driven methods through the example of simulating continuous-time systems from sampled data, and propose a pragmatic implicit behavioral description for polynomial systems, defined by the zero set of multiple, multivariate polynomials.
    \item We provide conditions on the data (akin to the generalized persistency of excitation for LTI systems) enabling a direct data representation of polynomial behaviors with bounded degree.
    This presents a distinct challenge compared to the LTI case, as not only the defining equations but also their polynomial multiples vanish on the data, so the relevant object is a polynomial ideal rather than a subspace.
    \item We propose computable data-driven simulation and predictive control methods based on the data representation, which are equivalent to their counterparts with known behavior under these conditions.
    These problems lead to polynomial systems of equations, which we solve globally with homotopy continuation (as in~\citet{rostalski2011numerical} for the model-based case).
    \item We illustrate the data-driven simulation and control methods on a Duffing oscillator, showing superior prediction accuracy compared to Koopman based schemes. Further, we confirm the benefits of implicit descriptions over explicit models numerically on a Van der Pol oscillator.
\end{itemize}

\textbf{Outline:} Section~\ref{sec:poly_behav} motivates implicit models and defines polynomial behaviors.
Section~\ref{sec:sysID} derives the direct data representation, which Section~\ref{sec:dd_methods} uses for data-driven simulation and control.
Section~\ref{sec:numerics} contains the numerical case studies, and Section~\ref{sec:conclusion} concludes the paper.

\textbf{Notation:}
let $\mathbb{R}$, $\mathbb{Z}$, and $\mathbb{N}$ denote the sets of real numbers, integers, and positive integers, respectively.
For a matrix $M\in\mathbb{R}^{m\times n}$, we denote its rank, image and null space by $\mathrm{rank}(M)$, $\mathrm{im}(M)$ and $\mathrm{null}(M)$, respectively.
Let $\mathbb{R}[x]$ denote the polynomial ring in $n\geq1$ variables $x$, and $\mathbb{R}[x]_{\leq d} \subset \mathbb{R}[x]$ the polynomials of degree at most $d$.
An \textit{ideal} $I \subset \mathbb{R}[x]$ is a set of polynomials that satisfies $0\in I$, $a+b\in I, -a \in I$, and $a\cdot c \in I$ for all $a,b\in I$ and all $c\in\mathbb{R}[x]$.
For an ideal $I$ of $\mathbb{R}[x]$, let $I_{\leq d} = I\cap\mathbb{R}[x]_{\leq d}$.
Given polynomials $p_i\in\mathbb{R}[x],~i\in\{1,\dots,r\}$, we denote the ideal generated by them as 
$$
\langle p_1,\dots,p_{r}\rangle = \left\{\sum_{i=1}^r c_i p_i ~\Big |~c_i \in \mathbb{R}[x]\right\}.$$
For any set $V$ in $\mathbb{R}^n$, $\mathcal{I}(V)$ denotes the \textit{vanishing ideal} of $\mathbb{R}[x]$ consisting of all polynomials that vanish on $V$, i.e., $
\mathcal{I}(V) = \{p\in\mathbb{R}[x]~|~p(x)=0,~\forall x \in V\}.
$
Conversely, for any set of polynomials $P$ in $\mathbb{R}[x]$, $\mathcal{Z}(P)$ denotes the common zeros of $P$, i.e., 
$
\mathcal{Z}(P) = \{x\in \mathbb{R}^n~|~p(x) =0,~\forall p\in P\}.
$

\section{Polynomial behaviors} \label{sec:poly_behav}
In this paper, we consider simulation and control of discrete-time nonlinear systems. To do this, we will first define the class of systems we consider in a precise manner. We approach this problem geometrically, which will allow us to capture important effects by encompassing implicit models, naturally generalize the linear case, and will enable a range of computationally feasible simulation and control methods.

\subsection{On implicit models} \label{sec:integration}
Physical systems are typically described by continuous-time dynamics, for example 
\begin{align} \label{eq:ode}
    \dot{x} = f(x),
\end{align}
for autonomous systems, where $x$ is the state and $f$ is a nonlinear function.
The function $f$ often takes a simple, e.g., polynomial or trigonometric form in practical applications.

Data-driven methods, on the other hand, work with discrete data
sampled every $T_s$ seconds, i.e., $x_k := x(kT_s)$.
Data-driven simulation aims at identifying the discrete relations $x_{k+1} = S_{T_s}(x_k)$, where $S_{T_s}$ is the solution map of~\eqref{eq:ode} over one sampling period.
However, $S_{T_s}$ rarely has a closed form and is typically more complex than $f$ itself.
Indeed, approximating $S_{T_s}$ is non-trivial even when $f$ is known, which is the subject of numerical integration.
Data-driven simulation and numerical integration (with a fixed step size) thus address the same question in a data-based and a model-based setting, respectively, and we draw on the latter to guide the design of the former.

Numerical integration methods are typically categorized as explicit or implicit, taking the form
\begin{align*}
    x_{k+1} = F_{T_s}(x_k) \qquad \text{or} \qquad G_{T_s}(x_k, x_{k+1}) = 0,
\end{align*}
respectively.
Although implicit methods are computationally more expensive per step because finding $x_{k+1}$ requires solving a nonlinear equation, their superior stability often permits much larger step sizes. 
This is crucial for stiff systems, where explicit methods require prohibitively small steps for stability, while implicit methods allow step sizes based purely on accuracy~\citep{hairer1996solving}.
The reason for this is that the implicit model class is strictly larger than the explicit (one can always pick $G_{T_s}(x_k, x_{k+1}) = x_{k+1} - F_{T_s}(x_k)$).
For example, for polynomial $f$, implicit integration schemes like the implicit Euler or trapezoidal rules induce the richer, algebraic explicit map required for stability, whereas explicit methods merely produce a polynomial map $F_{T_s}$.

Motivated by these observations, instead of explicit methods based on, e.g., nonlinear ARX models, we adopt the behavioral point of view that treats all variables symmetrically and defines a system by implicit relations between these variables.
The numerical case study in Section~\ref{sec:VdP} validates this choice, demonstrating that an implicit data-driven behavioral method successfully simulates a stiff system where an explicit model trained on the same data fails.

\subsection{Generalizing linear systems} 
In contrast to starting from a nonlinear set of defining difference equations, the behavioral approach views systems as subsets of the space of all trajectories. More specifically, the vast majority of methods consider only linear systems, which correspond to subspaces, or equivalently, to the zero set of linear equations. From the perspective of real algebraic geometry, the natural generalization of this is to consider systems described by zero sets of higher-order equations: \textit{real algebraic varieties}.

In mathematical terms, we consider signals $w: \mathbb{Z} \rightarrow \mathbb{R}^q$, which are $q$-variate signals collecting all variables of interest such as inputs, outputs, or states of the system.
The behavioral setting imposes no a priori input/output partition and treats all variables symmetrically, which makes it well suited for the modeling of implicit relations in the data, see Section~\ref{sec:integration}. 
In the behavioral setting a discrete-time system is identified with a set of possible trajectories $\mathcal{B} \subseteq \{w ~|~  w: \mathbb{Z} \rightarrow \mathbb{R}^q \}$. 
In this paper, we consider complete systems with lag $\ell$~\citep[cf.][]{willems1991paradigms}, which are also time-invariant, that is, for any valid trajectory of $\mathcal{B}$, any time-shift of the trajectory is also in $\mathcal{B}$.

To enable data-based methods, we need to make further assumptions on the set of allowed systems, or \textit{model class}. For this, we first define the \textit{restricted behavior} as
\begin{align} \label{eq:def_BL}
	\mathcal{B}_L:=\left\{\mathbf{w}\in\mathbb{R}^{qL}~|~\mathbf{w} = \begin{bmatrix}
		w_1^\top& w_{2}^\top &\dots & w_{L}^\top
	\end{bmatrix}^\top \textrm{ for some } w\in \mathcal{B} \right\}.
\end{align}
For time-invariant systems, this is the set of all length-$L$ windows of trajectories  $w\in\mathcal{B}$.
As our main goal is to propose computable data-driven simulation and control methods, we now define polynomial systems through their restricted behaviors \citep[cf.][Proposition~II.6]{willems1991paradigms}.
Instead of assuming that the system is linear, i.e., that $\mathcal{B}$ is a subspace, we will consider systems defined by the zero set of a set of polynomials, i.e., an algebraic variety, as formalized below.

\begin{ass} \label{ass:poly_behav}
	For any $L \geq \ell + 1$, there exists a number $n_p\in\mathbb{N}$ and a set of multivariate finite degree polynomials $p_i \in \mathbb{R}[\mathbf{w}]$, such that $\mathcal{B}_L$ is the zero set
	\begin{align} \label{eq:def_poly_behav}
		\mathcal{B}_L = \left\{\mathbf{w}\in\mathbb{R}^{qL}~|~p_i(\mathbf{w}) = 0,\quad \forall i\in\{1,\dots,n_p\} \right\}.
	\end{align}
\end{ass}
We then call $\mathcal{B}$ a polynomial behavior.
Note that, while an arbitrary set of polynomials can define a behavior $\mathcal{B}$ (see Appendix~\ref{app:behaviors}), not every such set constitutes a valid restricted behavior as in Assumption~\ref{ass:poly_behav}.
Characterizing whether a given set yields a well-posed system, such as having feasible initial conditions, unique solutions, causality, or even admitting real trajectories, is beyond the scope of the present paper.

The behavioral description~\eqref{eq:def_poly_behav} includes linear time-invariant systems, bilinear systems, as well as nonlinear input/output models with polynomial nonlinearities.
Moreover, since $w$ may also contain the state of the system, any state-space system with polynomial dynamics and output equation falls into the class of systems defined in this way.
In technical terms, the generating set~\eqref{eq:def_poly_behav} is a \textit{real algebraic variety}.
Next, we focus on deriving a direct data representation for $\mathcal{B}_L$ from Assumption~\ref{ass:poly_behav} and applying it to data-driven simulation and control.
To keep the presentation self-contained, we introduce only the algebraic geometry concepts necessary for this task, referring the reader to Appendix~\ref{app:AG} for additional details.

\section{Direct data representation} \label{sec:sysID}
Inspired by behavioral systems theory for LTI systems, we now investigate how the restricted polynomial behavior in~\eqref{eq:def_poly_behav} can be represented directly by data.
We choose a \textit{dictionary} $\phi(\mathbf{w}) = \begin{bmatrix}
    \phi_1(\mathbf{w}) & \dots & \phi_{n_\theta}(\mathbf{w})
\end{bmatrix}^\top$ consisting of polynomials $\phi_i \in \mathbb{R}[\mathbf{w}]_{\leq d_\mathrm{dict}}$ up to degree $d_\mathrm{dict}$.
Assume we measure $D$ data trajectories $\mathbf{w}^i\in\mathcal{B}_L,~i\in\{1,\dots,D\}$ (either windows of a single trajectory, or separate trajectories) and construct the lifted data matrix
\begin{align*}
    \mathbf{\Phi} := \begin{bmatrix}
        \phi(\mathbf{w}^1) & \phi(\mathbf{w}^2) & \dots & \phi(\mathbf{w}^D)
    \end{bmatrix} \in \mathbb{R}^{n_\theta \times D}.
\end{align*} 
For LTI systems, the dictionary contains only linear functions, so that $\mathbf{\Phi}$ is a trajectory (or Hankel) matrix.
For sufficiently rich data, its image is the restricted LTI behavior, while a suitable basis of $\mathrm{null}(\mathbf{\Phi}^\top)$ provides the coefficients of an ARX model (see Example~\ref{ex:LTI}).
We now focus on the conditions guaranteeing that $\mathbf{\Phi}$ provides a direct data representation of the polynomial behavior $\mathcal{B}_L$.

\begin{prb} \label{prb:sysID}
    Given data $\{\mathbf{w}^i\}_{i\in\{1,\dots,D\}}$ from $\mathcal{B}_L$ defined as in Assumption~\ref{ass:poly_behav}, find conditions on the dictionary and the data such that any $\mathbf{w}\in\mathcal{B}_L$ can be expressed in terms of $\mathbf{\Phi}$.
\end{prb}
The solution for this problem is well established for LTI behaviors~\citep{willems2005note}.
However, a new difficulty arises when polynomial behaviors are considered.
Namely, in addition to the defining polynomials $p_i$ in~\eqref{eq:def_poly_behav}, all polynomials in the form $\sum_{i}c_i p_i$ with $c_i\in\mathbb{R}[\mathbf{w}]$ vanish on the data $\mathbf{w}^i \in \mathcal{B}_L$, cf. Example~\ref{ex:duffing_2} below.
Therefore, we provide a direct data representation for $\mathcal{B}_L$ through the vanishing ideal $\mathcal{I}(\mathcal{B}_L)$ instead of the defining polynomials $p_i$.
In case the defining polynomials $p_i$ are given, the vanishing ideal can be expressed using the real radical, see Appendix~\ref{app:AG} for details.

\subsection{Conditions on the dictionary}
One could naively think that the dictionary degree $d_\mathrm{dict}$ has to be at least as large as the maximal degree of $p_i$ defining $\mathcal{B}_L$ in~\eqref{eq:def_poly_behav}.
However, this is not necessary, as we illustrate on an example below.
Instead, we define the \textit{minimal degree} of a polynomial behavior as
$$
\delta(\mathcal{B}_L) = \min \{d\in\mathbb{N} ~|~ \langle \mathcal{I}(\mathcal{B}_L)_{\leq d}\rangle = \mathcal{I}(\mathcal{B}_L)\}.
$$
In words, $\delta(\mathcal{B}_L)$ is the minimum degree, such that the polynomials with degree at most $\delta(\mathcal{B}_L)$ vanishing on $\mathcal{B}_L$ generate all polynomials vanishing on $\mathcal{B}_L$.
We illustrate the concept on an example.

\begin{example} \label{ex:duffing_1}
Consider the Duffing oscillator also used in~\citet{mezic2026tutorial,rostalski2011numerical}
\begin{align*}
    \dot{x}_1 = x_2, \qquad \dot{x}_2 = -\gamma x_2 - x_1(\alpha + \beta x_1^2) + u,
\end{align*}
with state $x=\begin{bmatrix} x_1 & x_2\end{bmatrix}^\top$ and input $u$.
The implicit trapezoidal rule with step $\Delta T$ yields the discrete-time dynamics defined by two polynomials $p_1(\mathbf{w})$ and $p_2(\mathbf{w})$ of degree 1 and 3 in the length-$2$ trajectory $\mathbf{w}$ with $w_t = \begin{bmatrix} u_t & x_{1,t} & x_{2,t}\end{bmatrix}^\top$.
Thus, $p_1$ and $p_2$ is a representation of $\mathcal{B}_2$ in~\eqref{eq:def_poly_behav}.
In this case, $\langle p_1,p_2\rangle = \mathcal{I}(\mathcal{B}_2)$ as the ideal $\langle p_1,p_2\rangle$ is real radical (cf. Appendix~\ref{app:AG}).
As no polynomials of smaller degree can generate $\langle p_1,p_2\rangle$, we have that $\delta(\mathcal{B}_2) = 3$.
In fact, $\delta(\mathcal{B}_L) = 3$ for any $L\geq 2$, as each time shift can be described by $p_1$ and $p_2$.
Alternative representations of the same $\mathcal{B}_2$ include $\{p_1,p_2,p_1\cdot p_2\}$ or $p_1^2 + p_2^2$, which are of larger degrees, $4$ and $6$, respectively.
The real zero set of these polynomials coincide with that of $\{p_1,p_2\}$, but their degree is not minimal.
\end{example}

To formulate the condition on the dictionary, note that $\mathbb{R}[\mathbf{w}]_{\leq d_{\mathrm{dict}}}$ is a linear subspace of $\mathbb{R}[\mathbf{w}]$ and the dimension is the number of monomials: $\mathrm{dim}(\mathbb{R}[\mathbf{w}]_{\leq d_{\mathrm{dict}}}) = \left(\begin{smallmatrix}
    qL + d_{\mathrm{dict}} \\ d_{\mathrm{dict}}
\end{smallmatrix}\right)$.
Further, any ideal $I\subseteq\mathbb{R}[\mathbf{w}]$ and set $I_{\leq d_\mathrm{dict}}$ are linear subspaces, as they are closed under addition and multiplication by constants.

\begin{ass} \label{ass:dict}
    Suppose $d_{\mathrm{dict}}\geq \delta(\mathcal{B}_L)$, $n_\theta = \mathrm{dim}(\mathbb{R}[\mathbf{w}]_{\leq d_{\mathrm{dict}}})$, and let $\phi$ be a basis for $\mathbb{R}[\mathbf{w}]_{\leq d_\mathrm{dict}}$.
\end{ass}

Assumption~\ref{ass:dict} ensures that the considered dictionary is sufficiently rich to explain $\mathcal{B}_L$.
Then, the lifted data matrix $\mathbf{\Phi}$ describes $\mathcal{I}\left(\{\mathbf{w}^i\}_{i\in\{1,\dots,D\}}\right)_{\leq d_{\mathrm{dict}}}$, i.e., the set of all polynomials in $\mathbb{R}[\mathbf{w}]_{\leq d_\mathrm{dict}}$ that vanish on the data, as formalized below.

\begin{lemma} \label{lemma:dict}
    Let Assumptions~\ref{ass:poly_behav} and \ref{ass:dict} hold. 
    Then, $\{\theta^\top \phi ~|~\theta \in \mathrm{null}(\mathbf{\Phi}^\top)\} =\mathcal{I}\left(\{\mathbf{w}^i\}_{i\in\{1,\dots,D\}}\right)_{\leq d_{\mathrm{dict}}}$, and this subspace has dimension $n_\theta - \mathrm{rank}(\mathbf{\Phi})$.
    Furthermore, 
    $\mathcal{I}\left(\{\mathbf{w}^i\}_{i\in\{1,\dots,D\}}\right)_{\leq d_{\mathrm{dict}}} \supseteq \mathcal{I}(\mathcal{B}_L)_{\leq d_\mathrm{dict}}$.
\end{lemma}
\begin{proof}
Consider any $\theta \in \mathrm{null}(\mathbf{\Phi}^\top)$.
By definition, $\theta^\top \phi(\mathbf{w}^i) = 0$ for any $i\in\{1,\dots,D\}$. 
Since $\phi$ is a basis for $\mathbb{R}[\mathbf{w}]_{\leq d_\mathrm{dict}}$, $\theta$ can describe any polynomial in $\mathbb{R}[\mathbf{w}]_{\leq d_\mathrm{dict}}$, and thus, the claim holds.
Moreover, the set inclusion holds since $\mathbf{w}^i\in\mathcal{B}_L$ implies $p(\mathbf{w}^i) = 0$ for all $p\in\mathcal{I}(\mathcal{B}_L)$.
\end{proof}
Next, we investigate the condition on the data formulated in terms of subspace dimensions.

\subsection{Conditions on the data} 
As we established, the nullspace of $\mathbf{\Phi}^\top$ represents all polynomials in $\mathbb{R}[\mathbf{w}]_{\leq d_\mathrm{dict}}$ vanishing on the data, and (for noise-free data) these include any polynomial vanishing on $\mathcal{B}_L$.
However, there might be more polynomials that vanish on the data than on $\mathcal{B}_L$ when the data are not rich enough.
Therefore, the condition on the data is that the set $\mathcal{I}(\mathcal{B}_L)_{\leq d_{\mathrm{dict}}}$ equals the set represented by $\mathrm{null}(\mathbf{\Phi}^\top)$.
As the sets are linear subspaces of $\mathbb{R}[\mathbf{w}]_{\leq d_\mathrm{dict}}$, this condition can be expressed by their dimensions.

\begin{ass}\label{ass:PE}
    $\mathrm{rank}(\mathbf{\Phi}) = n_\theta - \dim\left(\mathcal{I}(\mathcal{B}_L)_{\leq d_{\mathrm{dict}}}\right)$.
\end{ass}

Assumption~\ref{ass:PE} can be written equivalently as $\dim(\mathrm{null}(\mathbf{\Phi}^\top)) = \dim\left(\mathcal{I}(\mathcal{B}_L)_{\leq d_{\mathrm{dict}}}\right)$, and implies that at least $D\geq n_\theta - \dim\left(\mathcal{I}(\mathcal{B}_L)_{\leq d_{\mathrm{dict}}}\right)$ trajectories are required.
The number $n_\theta -\dim\left(\mathcal{I}(\mathcal{B}_L)_{\leq d_{\mathrm{dict}}}\right)$ in Assumption~\ref{ass:PE} grows with $d_{\mathrm{dict}}$, so a richer dictionary requires more data.
In case the generators $p_1,\dots,p_{n_p}$ in~\eqref{eq:def_poly_behav} are known, $\dim\left(\mathcal{I}(\mathcal{B}_L)_{\leq d_{\mathrm{dict}}}\right)$ can be calculated by first computing the real radical of $\langle p_1,\dots,p_{n_p}\rangle$, e.g., with the algorithm of~\citet{becker1993computation}, and then a Gr\"obner basis of it, see Appendix~\ref{app:AG} and~\citet[Ch.~9, \S3]{cox2015ideals}.
Under Assumptions~\ref{ass:dict} and~\ref{ass:PE}, the behavior can be fully described by the lifted data matrix as formalized next.

\begin{lemma} \label{lemma:FL}
    Let Assumptions~\ref{ass:poly_behav}, \ref{ass:dict}, and~\ref{ass:PE} hold, and consider the lifted data matrix $\mathbf{\Phi}$.
    Then, a trajectory $\mathbf{w}\in\mathcal{B}_L$ if and only if $\phi(\mathbf{w}) \in \mathrm{im}(\mathbf{\Phi})$.
\end{lemma}
\begin{proof}
By Assumption~\ref{ass:PE}, the inclusion in Lemma~\ref{lemma:dict} is tight: $
\mathcal{I}(\mathcal{B}_L)_{\leq d_\mathrm{dict}} = \mathcal{I}(\{\mathbf{w}^i\}_{i\in\{1,\dots,D\}})_{\leq d_\mathrm{dict}}$.
If $\phi(\mathbf{w}) \in \mathrm{im}(\mathbf{\Phi})$, then $p(\mathbf{w})=0$ for all $p \in \mathcal{I}(\mathcal{B}_L)_{\leq d_{\mathrm{dict}}}$ by Lemma~\ref{lemma:dict}.
Thus, $\mathbf{w} \in \mathcal{Z}(\mathcal{I}(\mathcal{B}_L)_{\leq d_{\mathrm{dict}}})$ $ =  \mathcal{Z}(\langle \mathcal{I}(\mathcal{B}_L)_{\leq d_{\mathrm{dict}}}\rangle) = \mathcal{Z}(\mathcal{I}(\mathcal{B}_L)) \subseteq \mathcal{Z}(\langle p_1,\dots,p_{n_p}\rangle) = \mathcal{B}_L$, where the second equality holds as $d_{\mathrm{dict}} \geq \delta(\mathcal{B}_L)$, and the set inclusion holds as $\langle p_1,\dots,p_{n_p}\rangle \subseteq \mathcal{I}(\mathcal{B}_L)$.
Conversely, if $\mathbf{w}\in\mathcal{B}_L$, then $p(\mathbf{w})=0$ for all $p\in\mathcal{I}(\mathcal{B}_L)_{\leq d_{\mathrm{dict}}}$, which holds iff $\phi(\mathbf{w}) \in \mathrm{im}(\mathbf{\Phi})$ by Lemma~\ref{lemma:dict}.
\end{proof}

\begin{example} \label{ex:duffing_2}
Consider again the Duffing oscillator from Example~\ref{ex:duffing_1} over a two-step horizon.
For Assumption~\ref{ass:dict} to hold, we need $d_{\mathrm{dict}}\geq \delta(\mathcal{B}_2) = 3$. 
Let us pick the minimal degree $d_{\mathrm{dict}}=3$.
Then, the polynomials in the set $\mathcal{I}(\mathcal{B}_2)_{\leq 3} = \langle p_1,p_2\rangle_{\leq 3}$ are $p_2$ and $c\cdot p_1$ for any $c\in\mathbb{R}[\mathbf{w}]_{\leq 2}$ (including the constant).
As $\mathrm{dim}(\mathbb{R}[\mathbf{w}]_{\leq 2}) = \left(\begin{smallmatrix}
    6 + 2 \\ 2
\end{smallmatrix}\right) = 28$, we have that $\dim\left(\mathcal{I}(\mathcal{B}_2)_{\leq d_{\mathrm{dict}}}\right) = 29$, and Assumption~\ref{ass:PE} requires $\mathbf{\Phi}$ to be of rank $n_\theta - 29 = 55$, which is the maximal rank $\mathbf{\Phi}$ can have.
Importantly, the direct data representation of $\mathcal{B}_2$ contains 29 polynomials, instead of $p_1$ and $p_2$ only.
\end{example}

\begin{example}[Linear and affine systems] \label{ex:LTI}
The considered setting with $d_\mathrm{dict} = 1$ recovers affine systems, whose restricted behavior is an affine subspace~\citep{padoan2023data,padoan2025time}.
LTI behaviors can be recovered by removing the constant term from the dictionary, i.e., $\phi(\mathbf{w}) = \mathbf{w}$.
Then, $\mathbf{\Phi}$ becomes a trajectory or Hankel matrix.
A linear form $\theta^\top\mathbf{w}$ vanishes on $\mathcal{B}_L$ if and only if $\theta\in\mathcal{B}_L^\perp$, so the linear forms in $\mathcal{I}(\mathcal{B}_L)$ form a subspace of dimension $\dim(\mathcal{B}_L^\perp) = qL-mL-n_x$ for $L>\ell$, where $m$ is the number of inputs and $n_x$ is the order.
For instance, $\theta$ can be chosen as the coefficients of an ARX model.
Assumption~\ref{ass:PE} then reads $\mathrm{rank}(\mathbf{\Phi}) = mL+n_x$, which is the generalized persistency of excitation condition~\citep{markovsky2023identifiability}, and Lemma~\ref{lemma:FL} states that $\mathbf{w}\in\mathcal{B}_L$ if and only if $\mathbf{w}\in\mathrm{im}(\mathbf{\Phi})$, which is the Fundamental Lemma~\citep{willems2005note}.
\end{example}

\begin{rmk}[Noisy data] \label{rmk:noise}
If the data are noisy, $\mathbf{\Phi}$ typically has full rank, i.e., $\mathrm{im}(\mathbf{\Phi})=\mathbb{R}^{n_\theta}$.
We then estimate $\mathrm{im}(\mathbf{\Phi})$ by computing the SVD of $\mathbf{\Phi}$ and keeping the singular vectors of the $h$ largest singular values.
The orthogonal complement of this truncated space then gives $n_\theta - h$ polynomials that approximately vanish on the data.
Since the true rank in Assumption~\ref{ass:PE} is unknown in practice, $h$ is a tuning parameter.
The truncation is optimal if the noise is additive on $\mathbf{\Phi}$ with i.i.d. zero-mean Gaussian entries, as it yields the best rank-$h$ approximation of $\mathbf{\Phi}$~\citep{eckart1936approximation}.
However, noise on the trajectories $\mathbf{w}^i$ enters $\mathbf{\Phi}$ through the monomials, making the noise on $\mathbf{\Phi}$ structured and not zero-mean.
The SVD ignores this structure, which biases the estimate.
The adjusted least squares method~\citep{markovsky2004consistent,usevich2016adjusted} corrects this bias and yields a consistent estimate.
\end{rmk}

\section{Data-driven simulation and control} \label{sec:dd_methods}
Now we use the direct data representation of the restricted behavior in Lemma~\ref{lemma:FL} to formulate the data-driven simulation and control problems without identifying a parametric model first.

\subsection{Data-driven simulation} \label{sec:simul}
Simulation amounts to computing the response of a system to a given input and initial condition~\citep{markovsky2008data}.
To formalize this, we partition $w_t$ into inputs $u_t$ and outputs $y_t$, and $\mathbf{w}$ into an initial part $\mathbf{w}_{\mathrm{ini}}$ of length $T_{\mathrm{ini}}$ specifying the initial condition, and a future part of length $T_{\mathrm{fut}} = L - T_{\mathrm{ini}}$ with inputs $u_{\mathrm{fut}}$ and outputs $y_{\mathrm{fut}}$.
With this partitioning, we structure the entire trajectory as $\mathbf{w} = \begin{bmatrix}
    \mathbf{w}_{\mathrm{ini}}^\top & u_{\mathrm{fut}}^\top & y_{\mathrm{fut}}^\top
\end{bmatrix}^\top$, where a suitable row permutation is omitted.
Note that $y_t$ may also be the state of the system, in which case a single sample suffices to specify the initial condition, i.e., $T_{\mathrm{ini}}=1$.

\begin{prb}[Simulation] \label{prb:simul}
    Given $\mathcal{B}_L$, $\mathbf{w}_{\mathrm{ini}}$, and $u_{\mathrm{fut}}$, find $y_{\mathrm{fut}}$ such that $\mathbf{w}\in\mathcal{B}_L$.
\end{prb}
As $\mathcal{B}_L$ is unknown, we apply Lemma~\ref{lemma:FL} to express the condition $\mathbf{w} \in \mathcal{B}_L$ by a system of polynomial equations through a basis $\Theta$ of 
$\mathrm{null}(\mathbf{\Phi}^\top)$, leading to an equivalent data-driven simulation problem.

\begin{theorem}[Data-driven simulation] \label{thm:simul}
    Let Assumptions~\ref{ass:poly_behav},~\ref{ass:dict}, and~\ref{ass:PE} hold.
    Then, for any $\mathbf{w}_{\mathrm{ini}}$ and $u_{\mathrm{fut}}$, Problem~\ref{prb:simul} has the same solutions as the data-driven simulation problem
    \begin{align} \label{eq:dd_simul}
        \textup{find } \quad y_{\mathrm{fut}} \quad \mathrm{s.t.} \quad \Theta\phi(\mathbf{w}) = 0.
    \end{align}
\end{theorem}
\begin{proof}
$\Theta\phi(\mathbf{w}) = 0$ if and only if $\phi(\mathbf{w})\in\mathrm{im}(\mathbf{\Phi})$, which is equivalent to $\mathbf{w}\in\mathcal{B}_L$ by Lemma~\ref{lemma:FL}.
\end{proof}

Since $\mathcal{B}_L$ provides an implicit description of the dynamics, Problem~\ref{prb:simul}, and hence~\eqref{eq:dd_simul}, may have multiple solutions, similar to implicit numerical integrators in Section~\ref{sec:integration}.
The appropriate solution then has to be selected, e.g., based on physical meaning or consistency with the past trajectory.

\begin{rmk}[Image formulation with regularization] \label{rmk:g}
Another common method to deal with noise in data-driven control is using the image of the lifted data matrix (i.e. writing $\mathbf{\Phi}(\mathbf{w}^d) g = \phi(\mathbf{w})$ for some $g$ in this setting) and adding regularization.
With a 2-norm regularizer the problem is still polynomial but now both in $\mathbf{w}$ and $g$. 
Thus, the number of unknowns increases limiting practical applicability.
However, since $g$ enters quadratically, it can be eliminated in closed form as in~\citet{sasfi2025gaussian,sasfi2026soft}, which 
leads to a formulation with a cost $J$ that is quadratic in $\phi(\mathbf{w})$.
The problem can be solved through its KKT conditions, but these form a polynomial system of degree $2d-1$, which is more expensive to solve than the degree-$d$ kernel representation.
\end{rmk}

\subsection{Data-driven predictive control} \label{sec:dd_control}
In predictive control, a finite-horizon optimal control problem is solved at every time step, and only the first optimal input is applied to the system.
With the partitioning of Section~\ref{sec:simul}, $\mathbf{w}_{\mathrm{ini}}$ contains the $T_{\mathrm{ini}}$ most recent measurements, and the future inputs and outputs $\mathbf{w}_{\mathrm{fut}} = \begin{bmatrix} u_{\mathrm{fut}}^\top & y_{\mathrm{fut}}^\top \end{bmatrix}^\top$ are the decision variables.
Following~\citet{rostalski2011numerical}, we consider a polynomial cost $J(\mathbf{w}_{\mathrm{fut}})$ and polynomial inequality constraints $c(\mathbf{w}_{\mathrm{fut}})\leq 0$.
A typical choice is the quadratic tracking cost $J(\mathbf{w}_{\mathrm{fut}}) = \|y_{\mathrm{fut}}-y_{\mathrm{ref}}\|_Q^2 + \|u_{\mathrm{fut}}\|_R^2$ with reference $y_{\mathrm{ref}}$ and weights $Q\succeq 0$, $R\succ 0$.

\begin{prb}[Predictive control] \label{prb:control}
    Given $\mathcal{B}_L$ and $\mathbf{w}_{\mathrm{ini}}$, solve
    \begin{align*}
        \min_{\mathbf{w}_{\mathrm{fut}}} \; J(\mathbf{w}_{\mathrm{fut}}) \quad \mathrm{s.t.} \quad \mathbf{w}\in\mathcal{B}_L, \quad c(\mathbf{w}_{\mathrm{fut}})\leq 0.
    \end{align*}
\end{prb}
As in Section~\ref{sec:simul}, applying Lemma~\ref{lemma:FL} to Problem~\ref{prb:control} leads to an equivalent data-driven formulation.
\begin{theorem}[Data-driven predictive control] \label{thm:control}
    Let Assumptions~\ref{ass:poly_behav},~\ref{ass:dict}, and~\ref{ass:PE} hold.
    Then, for any $\mathbf{w}_{\mathrm{ini}}$, $J$, and $c$, Problem~\ref{prb:control} has the same solutions as the data-driven predictive control problem
    \begin{align} \label{eq:dd_control}
        \min_{\mathbf{w}_{\mathrm{fut}}} \; J(\mathbf{w}_{\mathrm{fut}}) \quad \mathrm{s.t.} \quad \Theta\phi(\mathbf{w}) = 0, \quad c(\mathbf{w}_{\mathrm{fut}})\leq 0.
    \end{align}
\end{theorem}
\begin{proof}
	Similarly to Theorem~\ref{thm:simul}, the statement is a direct consequence of Lemma~\ref{lemma:FL}.
\end{proof}

\subsection{Solving the simulation and control problems}
The data-driven simulation problem~\eqref{eq:dd_simul} is a system of polynomial equations in $y_{\mathrm{fut}}$, which can be solved approximately with Newton-type methods or, for smaller systems, to global optimality with homotopy continuation~\citep{bates2013numerically}.
The data-driven control problem~\eqref{eq:dd_control} is a non-convex polynomial optimization problem, which can be solved locally, e.g., by sequential quadratic programming.
Instead, we reformulate~\eqref{eq:dd_control} as a system of polynomial equations via its KKT conditions, as done by~\citet{rostalski2011numerical} in the model-based case.
Under suitable constraint qualifications, every local minimizer satisfies
\begin{align} \label{eq:KKT}
\begin{split}
    &\nabla J(\mathbf{w}_{\mathrm{fut}}) + \partial\phi(\mathbf{w})^\top \Theta^\top \lambda + \partial c(\mathbf{w}_{\mathrm{fut}})^\top \nu = 0,\\
    &\Theta\phi(\mathbf{w}) = 0, \quad \nu_i\, c_i(\mathbf{w}_{\mathrm{fut}}) = 0~~\forall i, \quad \nu \geq 0, \quad c(\mathbf{w}_{\mathrm{fut}}) \leq 0,
\end{split}
\end{align}
for some multipliers $\lambda$ and $\nu$, where $\nabla$ and $\partial$ denote the gradient and the Jacobian with respect to $\mathbf{w}_{\mathrm{fut}}$.
Among the real solutions of~\eqref{eq:KKT}, the one with the lowest cost is the global minimizer.
Since $\mathbf{w}_{\mathrm{ini}}$ and $y_{\mathrm{ref}}$ enter~\eqref{eq:KKT} only as parameters, the system can be solved offline for generic parameter values, and only the paths starting at these solutions need to be tracked online~\citep{rostalski2011numerical}.
Note, however, that the computational cost of homotopy continuation grows exponentially with the number of unknowns, making it intractable for large horizons.
As a remedy, one can apply the data representation recursively, with the minimal $L = \ell + 1$, so that the computational cost grows linearly with the horizon.

\begin{rmk}[Homotopy continuation with redundant constraints] \label{rmk:HC}
Homotopy continuation computes all isolated solutions of a square polynomial system, i.e., one with as many equations as unknowns~\citep{bates2013numerically}.
Since the constraints $\Theta\phi(\mathbf{w})=0$ are highly redundant, problem~\eqref{eq:dd_simul} is overdetermined, and the solutions of~\eqref{eq:KKT} are not isolated, as the multipliers $\lambda$ are not unique.
Homotopy continuation methods typically resolve this issue by replacing $\Theta$ by $A\Theta$, where $A$ is a randomly chosen matrix that makes the system square.
Every solution of $\Theta\phi(\mathbf{w})=0$ also solves $A\Theta\phi(\mathbf{w})=0$, but one needs to check the feasibility of the computed solutions on the original constraint $\Theta\phi(\mathbf{w})=0$.
Instead of a random matrix, it is often preferable to minimize $\|\Theta \phi(\mathbf{w})\|^2$ rather than seeking an exact solution (see Remarks~\ref{rmk:noise} and~\ref{rmk:g}).
A practical choice for $A$ is then the Jacobian $\partial (\Theta \phi(\mathbf{w}))^\top$, as $2 A \Theta \phi(\mathbf{w}) = 0$ for the minimizer.
The Jacobian can be evaluated at a relevant reference point, such as the most recent measurement of $\mathbf{w}$ or the control target.
\end{rmk}

\section{Numerical case studies} \label{sec:numerics}
We illustrate the proposed methods on two benchmark nonlinear systems below.
All computations were carried out using Matlab, on a Lenovo ThinkPad X1 Carbon Gen 10 laptop with 12th Gen Intel i7-1260P × 16 CPU running Ubuntu 22.04.1 LTS.
The polynomial systems are solved with Bertini~\citep{bertini}.
As outlined in Remark~\ref{rmk:HC}, we square the system using the Jacobian evaluated at the current state or the origin for control, and retain the solution with the smallest residual.
The code reproducing the results is available at \url{https://gitlab.ethz.ch/asasfi/polynomial_behaviors}.

\subsection{Van der Pol oscillator} \label{sec:VdP}
We consider the Van der Pol oscillator
$$
\dot{x}_1 = x_2,\qquad \dot{x}_2 = \mu(1-x_1^2)x_2 - x_1,
$$
where increasing the parameter $\mu$ increases the stiffness of the system.
We assume that the system is unknown and that its state $w_k = \begin{bmatrix} x_1(kT_s) & x_2(kT_s)\end{bmatrix}^\top$ is measured with sampling time $T_s = 0.1$.
We use windows of length $L=2$ and $n_\theta=35$ monomials up to degree $d_{\mathrm{dict}}=3$.
The data are $6$ noise-free trajectories of $1000$ samples from initial states $(-3,-10), (0,-10), (3,-10), (-3,10),(0,10),(3,10)$ simulated with the stiff solver, \texttt{ode15s}, which is considered to be the true continuous-time trajectory.

Since the true trajectory does not satisfy polynomial relations up to degree $3$ exactly, we compute $\Theta$ from the truncated SVD of $\mathbf{\Phi}$ (Remark~\ref{rmk:noise}).
The rank is selected manually as $h=32$, since there is a visible drop in the singular values of $\mathbf{\Phi}$ at that value.
The prediction is then computed recursively by solving~\eqref{eq:dd_simul} starting always from the last predicted state.
As a benchmark, we identify an explicit model $w_{k+1} \approx \Xi \phi(w_k)$ (cf. Section~\ref{sec:integration}), where $\Xi$ minimizes the sum of $\|w_{k+1} - \Xi \phi(w_k)\|_2^2$ over the $D$ state pairs $(w_k, w_{k+1})$ used to build $\mathbf{\Phi}$.
The behavioral method takes about 20 ms per prediction step, whereas the explicit model takes less than 0.1 ms.

\begin{figure}
    \centering
    \includegraphics[width=0.49\linewidth]{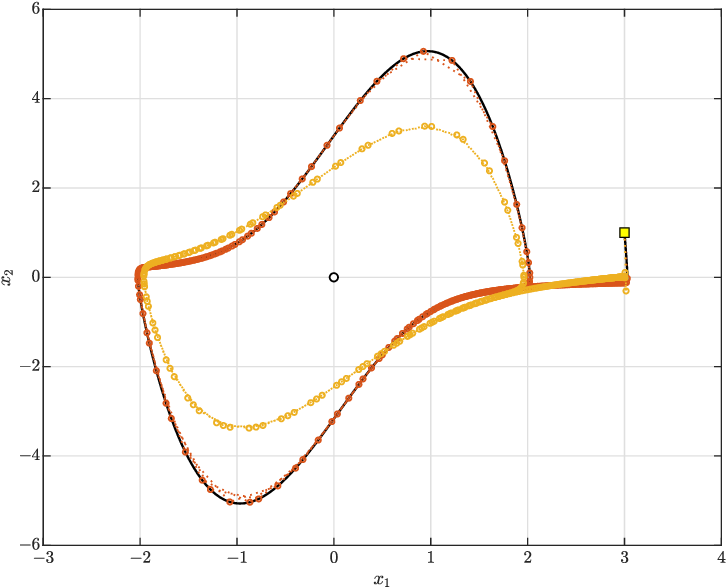}
    \includegraphics[width=0.49\linewidth]{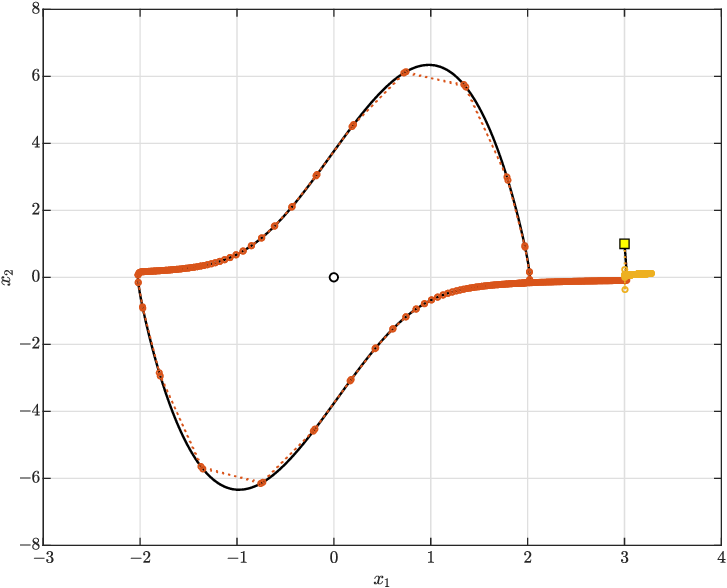}
    \caption{Van der Pol oscillator trajectories for $\mu=3$ (left) and $\mu=4$ (right), showing predictions over 300 steps starting from the initial state $w_0$ (yellow square). The true trajectory is the black line, the behavioral method is in orange, and the explicit benchmark is in yellow.}
    \label{fig:VdP4}
\end{figure}

We test the methods on a simulation task for $300$ steps starting from the initial condition $w_0 = \begin{bmatrix} 3 & 1\end{bmatrix}^\top$.
As shown in Figure~\ref{fig:VdP4}, the behavioral method stays close to the true trajectory throughout the prediction for both $\mu=3$ and $\mu=4$.
In contrast, the explicit model only captures the qualitative shape of the limit cycle for $\mu=3$ with a significant offset, and does not produce a meaningful prediction for the stiffer system with $\mu=4$.
Note that the explicit Euler and Runge-Kutta 4 numerical integration methods on the same task become unstable for $\mu>2$ and $\mu>3$, respectively.

\subsection{Duffing oscillator} \label{sec:Duffing}
Next, we consider the forced Duffing oscillator from~\citet{mezic2026tutorial}
$$
\dot{x}_1 = x_2,\qquad \dot{x}_2 = -\gamma x_2 - \alpha x_1 - \beta x_1^3 + u,
$$
with the parameters $\gamma=0.5$, $\alpha=-1$ and $\beta=1$, see Figure~\ref{fig:Duffing_control} for the phase portrait.
The input and the state are measured with sampling time $T_s=0.05$, and we use zero-order hold between the samples for $u$.
We adopt the code accompanying \citet{mezic2026tutorial} and use the data from their example to construct $\mathbf{\Phi}$.
As for the Van der Pol oscillator, no polynomial relation holds exactly between the samples, and we compute $\Theta$ from the truncated SVD of $\mathbf{\Phi}$ (Remark~\ref{rmk:noise}) in all experiments, also for noise-free data.

\begin{figure}
    \centering
    \includegraphics[width=0.49\linewidth]{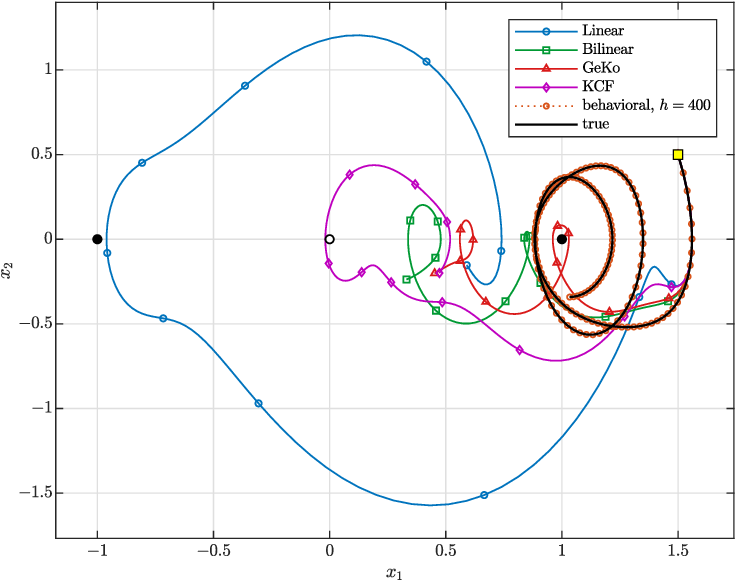}
    \includegraphics[width=0.49\linewidth]{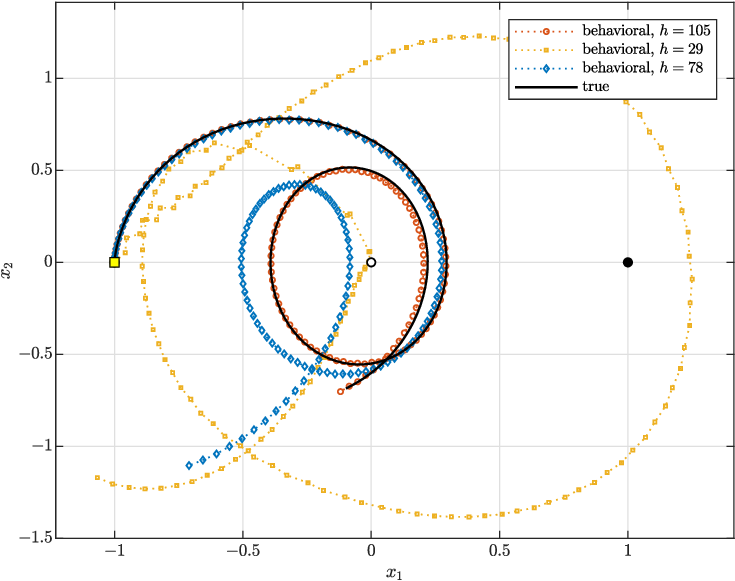}
    \caption{Predictions for the Duffing oscillator starting from initial state $x_0$ (yellow square). 
    Black lines denote the true trajectory, while markers connected by dotted lines represent the behavioral method.
    Left: Exact data test case from~\citet{mezic2026tutorial}, comparing the behavioral method (orange) against different Koopman models with polynomial lift trained on the same data (linear -- blue, bilinear -- green, GeKo -- red, and KCF -- magenta).
    Right: Predictions under noise using the behavioral method with different truncations $h=105$ (orange circles), $h=29$ (yellow squares), and $h=78$ (blue diamonds).}
    \label{fig:Duffing_sim}
\end{figure}

\subsubsection{Simulation} \label{sec:Duffing_sim}
We set $L=3$ with $T_{\mathrm{ini}}=1$ and degree $d_{\mathrm{dict}}=4$ (matching the polynomial lift of~\citet{mezic2026tutorial}) giving $n_\theta=715$ .
The singular values of $\mathbf{\Phi}$ decay without a visible gap, but the predictions are not sensitive to the choice of $h$ for a large interval in the noise-free case, and we pick $h=400$ from this interval.
The 150 steps long test trajectory from~\citet{mezic2026tutorial} is predicted by solving~\eqref{eq:dd_simul} recursively, starting from the last predicted state (75 solves).
The predictions of various Koopman modelsfrom~\citet{mezic2026tutorial}, trained on the same data, are shown in Figure~\ref{fig:Duffing_sim} (left).
Our prediction overlaps with the true trajectory, whereas the Koopman predictions drift away.
This is likely because the Koopman models propagate the lifted state $z$ linearly, without enforcing that $z$ remains in the range of the lifting.
In contrast, our method solves the identified implicit polynomial relations directly, at the price of increased computational load.

Next, we add zero-mean Gaussian noise with standard deviation $10^{-3}$ to the state data and to the initial state.
We also use a more demanding test case, in which the input $u(t) = 1.25\sin(1.6t)$ drives the state from the stable equilibrium $x_0=(-1,0)$ to the other basin of attraction.
Figure~\ref{fig:Duffing_sim} (right) shows the predictions of our method for three different truncated ranks.
The best truncation is at $h=105$ providing virtually zero error, whereas $h=78$ tracks the true trajectory with larger error.
The trajectory produced by truncation at $h=29$ attains a very large error, although the shape resembles that of the true trajectory.

\subsubsection{Control} \label{sec:Duffing_control}
\begin{figure}
    \centering
    \begin{minipage}[c]{0.5\textwidth}
    \centering
    \includegraphics[width=\linewidth]{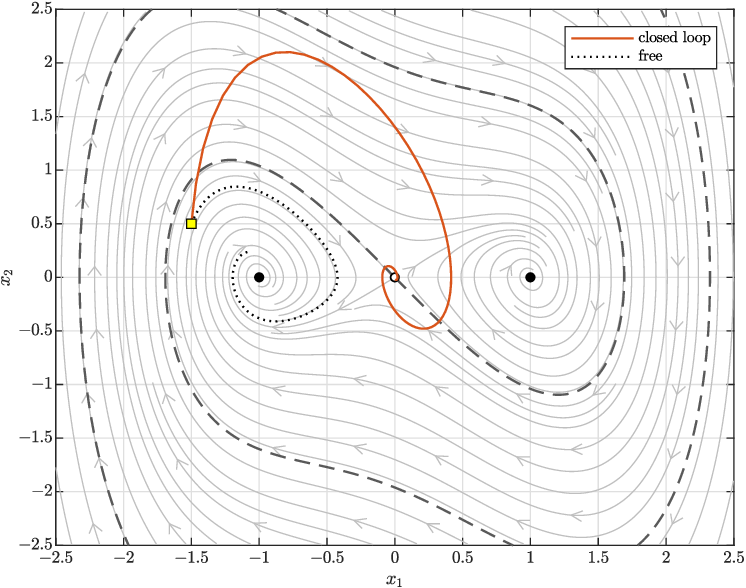}
    \end{minipage}\hfill
    \begin{minipage}[c]{0.48\textwidth}
    \raggedright
    \caption{Duffing oscillator under the proposed controller starting from $x_0$ (yellow square). Unlike the free response (dotted), the closed-loop trajectory (orange) crosses the dashed separatrix from the basin of $(-1,0)$ to stabilize at the saddle at $(0,0)$. The background shows the unforced phase portrait and the basins of attraction for the stable equilibria (filled circles).}
    \label{fig:Duffing_control}
    \end{minipage}
\end{figure}

We regulate the oscillator to the saddle at the origin with the data-driven predictive controller~\eqref{eq:dd_control}, using $Q=\mathrm{diag}(100,1)$ and $R=0.1$.
To keep the offline computation tractable, we reduce the degree to $d_{\mathrm{dict}}=3$ ($n_\theta=220$) and choose $h=150$.
The offline solve takes about 10 minutes, and the online computations about $0.07$\,s per time step.
Figure~\ref{fig:Duffing_control} shows the closed loop from $x_0=(-1.5,0.5)$.
Without control, the state converges to $(-1,0)$, but the controller steers the state across the separatrix to the other basin of attraction and then stabilizes at the origin.

\section{Conclusion} \label{sec:conclusion}
We extended direct data representations from LTI to polynomial behaviors by describing their finite-length trajectories as a real algebraic variety.
We provided conditions on the data so that the behavior has a direct data representation.
This implicit description leads to data-driven simulation and control problems that can be solved globally with homotopy continuation.
The proposed methods and their benefits over existing schemes were illustrated on benchmark nonlinear case studies.
Future work includes provable robustness to noisy data and solution methods that scale to long horizons.

\bibliography{Literature}

\appendix
\section{} \label{app:behaviors}
Here we detail the remark after Assumption~\ref{ass:poly_behav}, namely that an arbitrary set of polynomials defines a behavior, but need not describe its restricted behavior.
Let $L \geq \ell + 1$ and consider a generating set
\begin{align*}
	\mathcal{S}:=\left\{\mathbf{w}\in\mathbb{R}^{qL}~|~p_i(\mathbf{w}) = 0,\quad \forall i\in\{1,\dots,n_p\} \right\},
\end{align*}
where $p_i \in \mathbb{R}[\mathbf{w}]$ are multivariate finite degree polynomials.
Any such set defines a behavior as
\[ \mathcal{B} = \{ w ~|~\begin{bmatrix}
w_t^\top& w_{t+1}^\top &\dots & w_{t+L-1}^\top
\end{bmatrix}^\top  \in \mathcal{S}, \quad \forall t\in\mathbb{Z} \}.\]
Most notably, the generating set $\mathcal{S}$ is not unique, as different sets $\mathcal{S}$ can give rise to the same system.
By definition, the restricted behavior $\mathcal{B}_L$ in~\eqref{eq:def_BL} of this system satisfies $\mathcal{B}_L\subseteq \mathcal{S}$ for any generating set, but the inclusion can be strict, since $\mathcal{S}$ may contain windows that do not extend to a trajectory of $\mathcal{B}$.
If $\mathcal{B}_L$ is itself the zero set of polynomials, as required by Assumption~\ref{ass:poly_behav}, then $\mathcal{B}_L$ is the smallest generating set of $\mathcal{B}$ and thus a canonical choice.

\section{} \label{app:AG}
Let $I$ be an ideal of $\mathbb{R}[x]$.
The real radical of $I$ is defined as
$$
\sqrt[R]{I} = \left\{p\in\mathbb{R}[x]~|~\exists m\in\mathbb{N},~\exists b_1,\dots,b_r\in\mathbb{R}[x],~p^{2m} + b_1^2 + \dots + b_r^2 \in I\right\},
$$
and $I$ is called real radical if $\sqrt[R]{I} = I$.
Intuitively, $\sqrt[R]{I}$ collects every polynomial that is implied by the equations in $I$ on real points.
For example, consider a system described by two polynomials $p_1$ and $p_2$.
The three ideals $I_1 = \langle p_1,p_2\rangle$, $I_2=\langle p_1^2,p_2\rangle$, and $I_3 = \langle p_1^2+p_2^2\rangle$ are different, as $p_1\in I_1$, but $p_1 \notin I_2,I_3$, and $p_2 \in I_1,I_2$, but $p_2\notin I_3$.
On the other hand, they all have the same real radical, i.e., $\sqrt[R]{I_1} = \sqrt[R]{I_2} = \sqrt[R]{I_3} = I_1$.
Thus, $I_1$ is a real radical ideal, but $I_2$ and $I_3$ are not.

The real Nullstellensatz (see, e.g.,~\cite{marshall2008positive}) relates the real radical to the vanishing ideal.
\begin{lemma}[Real Nullstellensatz]
    Let $I$ be an ideal of $\mathbb{R}[x]$. Then, $\mathcal{I}(\mathcal{Z}(I)) = \sqrt[R]{I}$.
\end{lemma}
Applied to $I = \langle p_1,\dots,p_{n_p}\rangle$ from~\eqref{eq:def_poly_behav}, the lemma gives
\begin{align*}
    \mathcal{I}(\mathcal{B}_L) = \sqrt[R]{\langle p_1,\dots,p_{n_p}\rangle}.
\end{align*}
This result allows us to obtain the vanishing ideal from any set of polynomials representing $\mathcal{B}_L$ in~\eqref{eq:def_poly_behav} by computing the real radical.

\end{document}